\documentclass[pss,fleqn]{w-art}
\usepackage{times}
\usepackage{w-thm}
\usepackage[]{graphicx}
\begin{document}
\DOIsuffix{theDOIsuffix}
\Volume{XX}
\Issue{X}
\Month{XX}
\Year{2004}
\pagespan{3}{}
\Receiveddate{May 2004}
\Reviseddate{XXX}
\Accepteddate{XXX}
\Dateposted{XXX}
\keywords{}
\subjclass[pacs]{{05.10.Ln},{75.40.Mg},{75.50.Tt},{75.60.Lr}}



\title[]{Magnetic relaxation in a model of interacting nanoparticles in terms of microscopic energy barriers}


\author[\`O. Iglesias]{\`Oscar Iglesias \footnote{Corresponding
     author: e-mail: {\sf oscar@ffn.ub.es}, Phone: +34\,93\,4021155,
     Fax: +34\,93\,4021149}\inst{1}} \address[\inst{1}]{Departament de \ F\'{\i}sica Fonamental, Universitat de Barcelona, Diagonal 647, 08028 Barcelona, Spain}
\author[A. Labarta]{Am\'{\i}lcar Labarta\inst{1}}
\begin{abstract}
Monte Carlo simulations are used to study the magnetic relaxation of a system of single domain particles with dipolar interactions modeled by a chain of Heisenberg classical spins. We show that the so-called $T\ln(t/\tau_0)$ method can be extended to interacting systems and how, from the computed master relaxation curves, the effective energy barrier distributions responsible for the relaxation can be obtained. A transition from a quasi-logarithmic to power-law behavior of the relaxation as the interaction strength is in-creased is found. By direct computation of the effective energy barriers of the system, we show that this is due to the appearance of an increasing number of small energy barriers caused by the reduction of the anisotropy energy barriers as the local dipolar fields increase.
\end{abstract}
\maketitle                   





\section{Introduction}

Dipolar interactions are ubiquitous in most magnetic nanoparticle systems due to their long-range char-acter. Many of the peculiar or anomalous phenomena in this kind of systems are direct consequence of the predominance of dipolar interactions over exchange interparticle interactions. Current preparation techniques and materials synthesis allow to study the role of dipolar interactions on the magnetic proper-ties in a controlled manner. For example, in granular metal solids the variation of volume metal fraction and size of the clusters allows to tune the strength of interparticle interactions \cite{Sahooapl03}. A similar control can be achieved in ferrofluids by dilution of the magnetic particles embedded in it \cite{Troncjm03}. More recently, one-dimensional arrays of nanoelements have also been the focus of several studies \cite{Cowburnscience00}. In particular, there is current controversy on whether the slow relaxation and blocking phenomena observed in particle systems is to be ascribed to dipolar interactions or is indicative of a spin-glass-like state at the particle surface \cite{Batllejpd02}. Here we present the results of Monte Carlo (MC) simulations aiming to clarify the role played by dipolar interactions in the time dependent magnetic properties of nanoparticle systems, establishing a connection between the microscopy energy landscape of the magnetic system and the observed relaxation laws by means of the so-called $T\ln(t/\tau_0)$ scaling.

\section {Model and simulation method}
The model consists of a linear chain (along the $x$ axis) of $N$ classical Heisenberg spins $\bf{S}_i$ which are intended to represent the a monodomain particle with magnetic moment $\bf{\mu}_i = \mu \bf{S}_i$. The particle easy-axis directions $\bf{n}_i$ are randomly distributed and the anisotropy constants $K_i$ are distributed according to a log-normal distribution of width $\sigma = 0.5$ and mean value $K_0= 1$. Therefore, the energy of the system can be written as   		
\begin{eqnarray}
	\cal{H}= -\sum_{i=1}^{N}\lbrace K_i (\bf{S}_i\cdot \bf{H}_i ^{eff} )
	\rbrace
\label{Eq1}
\end{eqnarray}
where $\bf{H}_i^{eff} = \bf{H} + \bf{H}_i^{dip}$ is the effective field at site $i$, $\bf{H}$ is the external magnetic field and 
\begin{eqnarray}
{\bf H}^{dip}_i=-g\sum_{j\neq i}^{N} 
\left\{ \frac{{\bf S}_j}{r_{ij}^3}-3
\frac{({\bf S}_j \cdot {\bf r}_{ji}){\bf r}_{ij}}
{r_{ij}^5}\right\} \ 
\end{eqnarray}
is the dipolar field felt by spin $\bf{S}_i$, where $g= \mu_0\mu^2/4\pi a^3$ characterizes the strength of the dipolar interaction. Simulations of the relaxation of the magnetization have been performed by means of the MC method with standard Metropolis algorithm. As for the details, let us only mention that, in order to be able to simulate successfully the long-time relaxation of the magnetization, we consider trial jumps between spin orientations corresponding to energy minima. Therefore, the energy jumps appearing in the acceptance probability always correspond to real energy barriers, at difference with most implementations of the MC method for continuous spins. In order to be able to compute the exact energy minima and barriers from Eq. \ref{Eq1}, spins are restricted to point inside the x-z plane. In this way, we have been able build an algorithm that computes exactly the maxima and minima of the energy as well as the corresponding energy barriers.

\section{$T\ln(t/\tau_0)$ scaling and effective energy barrier distributions}

\begin{figure}[h]
\centering
\includegraphics[width= 0.5\columnwidth]{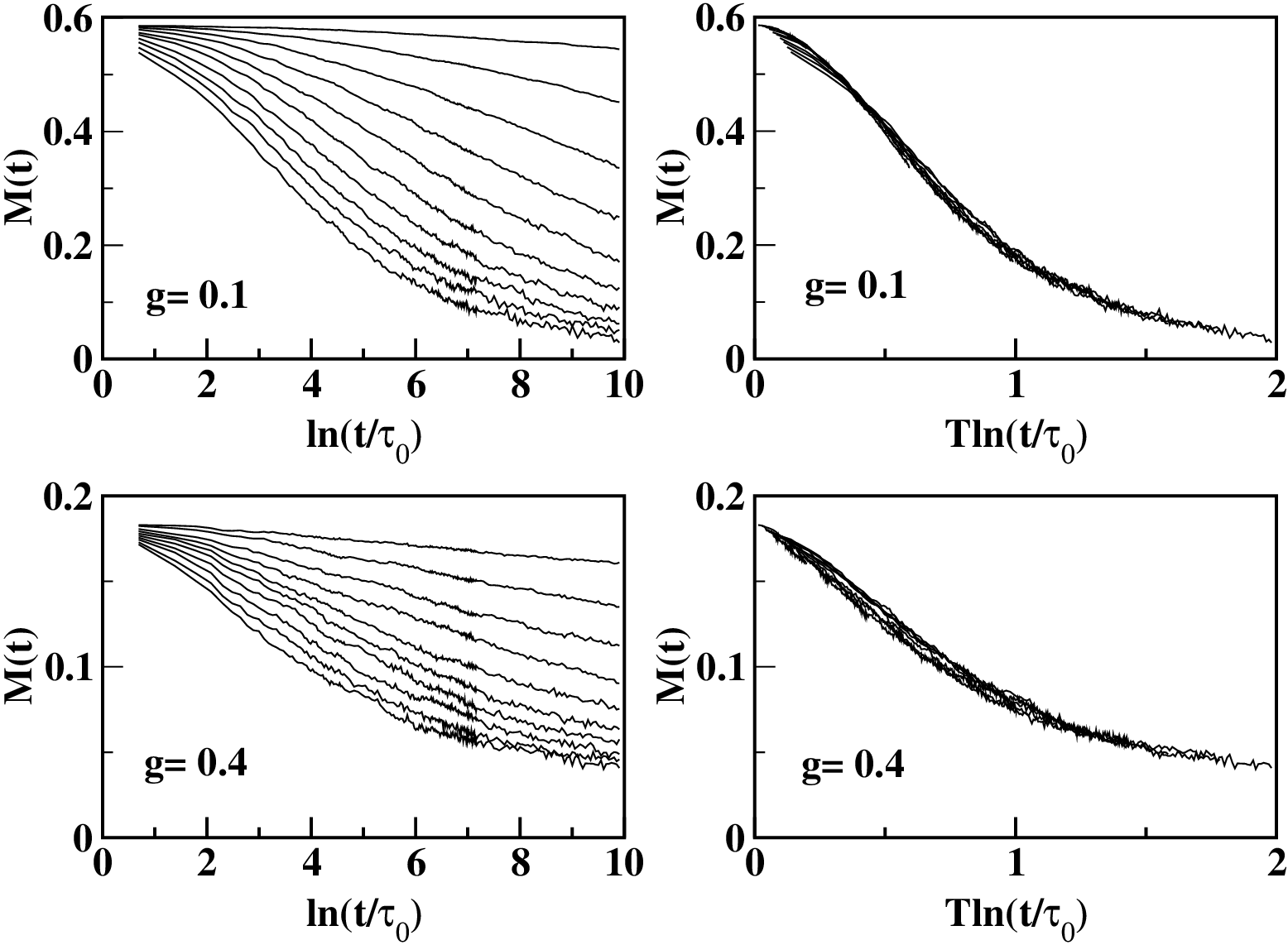}
\caption{Left panels: relaxation curves for temperatures ranging from $T= 0.02$ (uppermost) to $0.2$ (lowermost curves) in $0.02$ steps for dipolar interaction strength $g= 0.1, 0.4$. Right panels: master curves corresponding to the relaxations shown in the left panels obtained after scaling by the respective multiplicative factor $T$.
} 
\label{Fig_1}
\end{figure}

In order to simulate the relaxation curves, we start from a configuration in which the spins are aligned in parallel along the z axis, a situation which mimics the application of a saturating magnetic field. Before starting the simulation, the spins are submitted to a previous equilibration process at $T= 0$, during which they are consecutively reoriented along the nearest energy minima during a number of MC steps until the system reaches an equilibrated state. Starting from this state, the relaxation towards the equilibrium state in zero applied field and finite $T$ has been computed for values of $g$ between $0.1$ and $0.5$ during at least $10000$ MC steps. The obtained relaxation curves have been analyzed following the $T\ln(t/\tau_0)$ scaling method introduced in our previous works \cite{Labartaprb93,Iglesiaszpb96,Balcellsprb97} for non-interacting systems. The master relaxation curves obtained after scaling of the curves at different $T$ along the horizontal axis by multiplicative factors $T$ are presented in Fig. \ref{Fig_1} for a range of temperature covering one order of magnitude. 

\begin{figure}[h]
\centering
\includegraphics[width= 0.7\columnwidth]{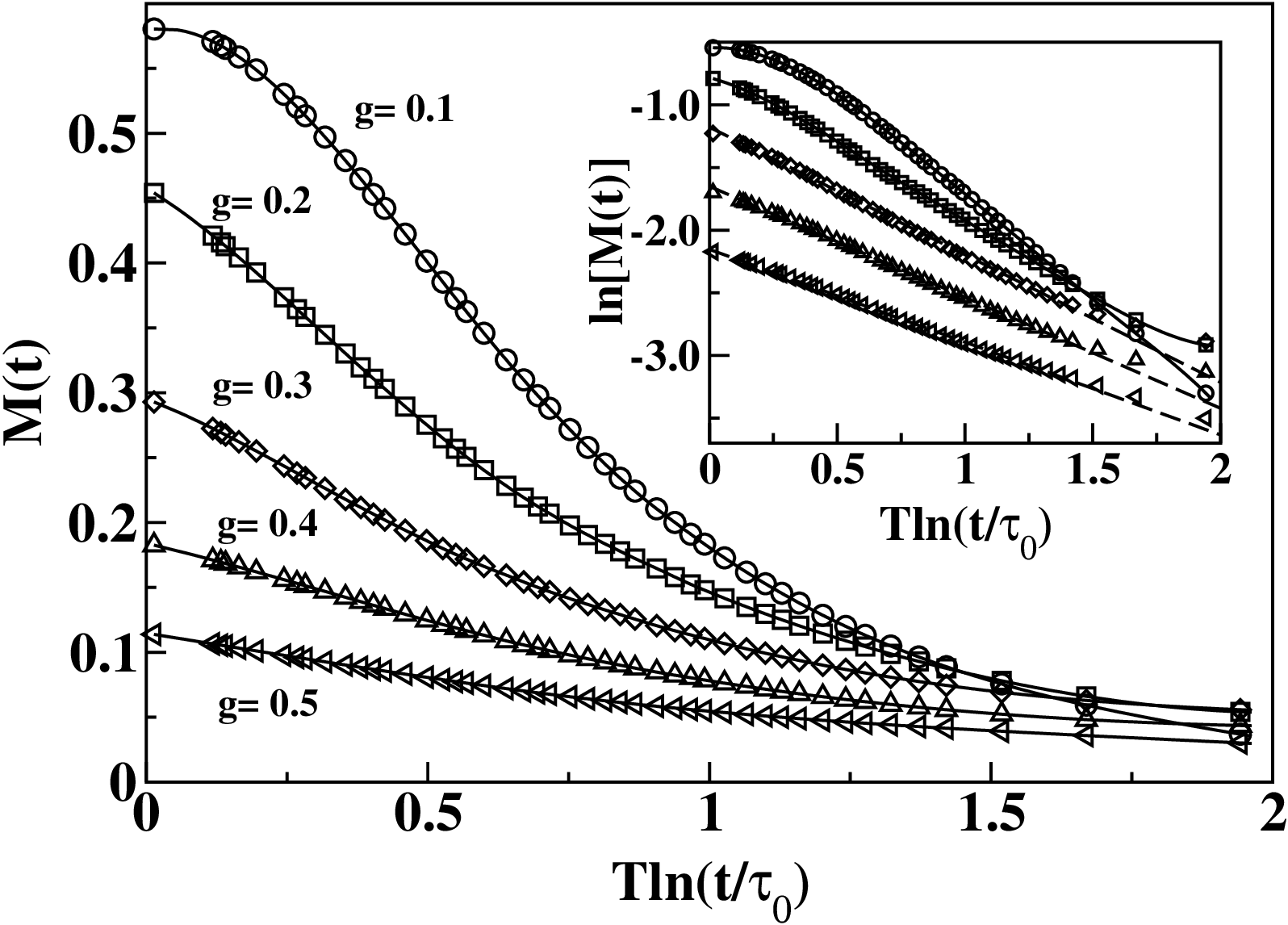}
\caption{Master relaxation curves for dipolar interaction strengths $g= 0.1, 0.2, 0.3, 0.4, 0.5$. The inset shows the same curves in a logarithmic scale together with the fittings to $m(t)\propto t^{-\gamma}$ (dashed lines).
} 
\label{Fig_2}
\end{figure}

Let us first notice that the curves overlap in a wide range of times and temperatures, this overlap being better for the lowest $T$ relaxations, while for the high $T$ curves it is accomplished only above the inflection point. This result supports the validity of the scaling in the presence of dipolar interactions, and gives support to experimental observations in systems where interactions cannot be neglected \cite{Luisprb02}. When dipolar interparticle interactions are considered, the energy barriers responsible for the thermal relaxation change as the relaxation proceeds but, as we have checked, this does not preclude the accomplishment of the $T\ln(t/\tau_0)$ scaling. In order to understand this, we have computed, within the scope of our model, the evolution of the distribution of energy barriers at different stages of the relaxation process and proved that, in spite of the dynamic change of the local energy barriers, they remain constant during the relaxation process. The second point studied has been the influence of the dipolar coupling strength $g$ on the relaxation law. For this purpose, we have plotted in Fig. \ref{Fig_2} the master relaxation curves for $g$ ranging from $0.1$ to $0.5$ after smoothing and filtering of the curves in Fig. \ref{Fig_1}. Inspection of the curves shows that there is a change in the relaxation law with increasing $g$. Whereas for weak interaction ($g= 0.1, 0.2$) the curves show an inflection point around which the decay is quasi-logarithmic, in the strong interaction regime ($g= 0.3, 0.4, 0.5$), the curves have downward curvature and can be fitted to a power-law decay of the form $m(t)\propto t^{-\gamma}$ with an exponent that decreases with increasing $g$, $\gamma =1.02, 0.80, 0.74$ for $g= 0.3, 0.4, 0.5$ (dashed lines in the inset of Fig. \ref{Fig_2}. This power law behavior has also been observed experimentally in arrays of micromagnetic dots \cite{Hyndmanjm02} and discontinuous multilayers \cite{Chenprb03} and also in other MC simulations of Ising spins \cite{Ribasjap96} and two dimensional spin systems \cite{Sampaioprb01}. 

\begin{figure}[htbp]
\centering
\includegraphics[width= 0.8\columnwidth]{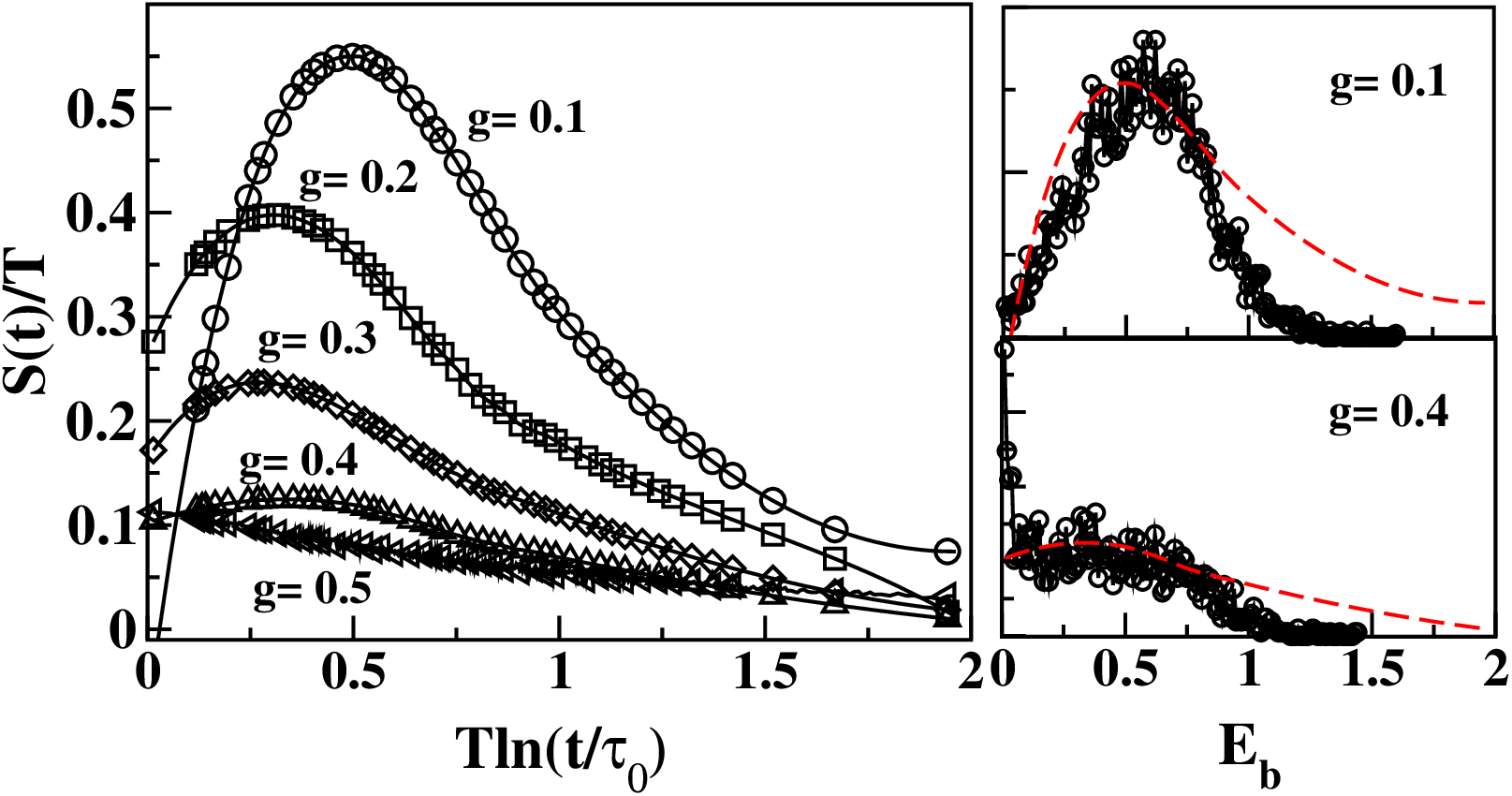}
\caption{Left panel: effective energy barrier distributions obtained form the derivatives of the master curves shown in Fig. \ref{Fig_2} for several values of the dipolar interaction strength $g$. Right panels: cumulative histograms of jumped energy barriers during a relaxation process at $T= 0.1$ after $t= 10000$ MC steps (symbols). Dashed lines stand for the derivatives of the master curves shown in the left panel.
} 
\label{Fig_3}
\end{figure}

In order to better understand the origin of this change in the relaxation laws, we will proceed to extract some information about the microscopic energy barriers responsible for the thermal relaxation in both regimes. For this purpose, we will use the master curves previously obtained to apply the method proposed previously for non-interacting systems \cite{Labartaprb93}. By performing the logarithmic time derivative of the master curves $S(t)= dM(t)/dln(t)$, it is possible to infer the effective energy barrier distribution feff(E) that would give rise to the relaxation curve corresponding to the master curve. Fig. \ref{Fig_3} presents the results for the same values of $g$ as in Fig. \ref{Fig_2}. The main features observed are the broadening of the energy barrier distribution and the increasing contribution of small energy barriers as the dipolar interaction $g$ increases.

For weak interaction values ($g= 0.1$), $f_{eff}(E)$ has a shape very similar to that for the non-interacting case, but its width increases with $g$ and the mean effective barrier corresponding to the maximum shifts to-wards lower energies. However, when entering the strong interaction regime (marked by the appearance of a contribution of almost zero energy barriers), the effect of the interaction no longer consists in a distortion of the original distribution. In this case, the interaction creates high energy barriers resulting in a more uniform $f_{eff}(E)$, that now becomes a decreasing function of the energy extending to higher values of the energy. In order to unveil the information given by the $f_{eff}(E)$ in the interacting case, we have also computed the cumulative histograms of energy barriers that have been jumped during the relaxation process simulated by the MC method. Results obtained after $10000$ MCS are shown for two values of $g$ in the left panels of Fig. \ref{Fig_3} and are compared to the distributions obtained from the master curves (dashed lines in the right panels of Fig. \ref{Fig_3}. If the histograms are computed by counting only the energy barriers that have not been already jumped up to a time $t$, we see that they coincide with the distributions deduced from the master relaxation curves. The differences between both results at high energy values are due to very high energy barriers that can only be surmounted at much longer times for the temperature considered here.
\begin{acknowledgement}
  We acknowledge CESCA and CEPBA under coordination of C$^4$ for computer facilities.
  This work has been supported by the spanish SEEUID through the MAT2003-01124 project
  and the Generalitat de Catalunya through the 2001SGR00066 CIRIT project.
\end{acknowledgement}

\end{document}